\documentclass{emulateapj}

\newcommand{\be}{\begin{equation}}
\newcommand{\ee}{\end{equation}}
\newcommand{\beqn}{\begin{eqnarray}}
\newcommand{\eeqn}{\end{eqnarray}}
\newcommand{\bld}[1]{\mbox{\boldmath$#1$\unboldmath}}
\newcommand{\gn}{{{G}}}
\newcommand{\C}{{{C}}}

\newcommand{\omn}{{{\eta}}}

\newcommand{\pomega}{{{\varpi}}}

\begin{document}

\title{The Effect of Charon's  Tidal Damping on the Orbits of Pluto's Three Moons}
\author{Yoram Lithwick\altaffilmark{1} \& Yanqin Wu\altaffilmark{2}}
\altaffiltext{1}{CITA, Toronto ON Canada}
\altaffiltext{2}{Dept. of Astronomy \& Astrophysics, University of Toronto, Toronto ON Canada}
\begin{abstract}

Pluto's recently discovered minor moons, Nix and Hydra, have almost
circular orbits, and are nearly coplanar with Charon, Pluto's major
moon.  This is surprising because tidal interactions with Pluto are
too weak to damp their eccentricities.  We consider an alternative
possibility: that Nix and Hydra circularize their orbits by
exciting Charon's eccentricity via secular interactions, and Charon in
turn damps its own eccentricity by tidal interaction with Pluto.  The
timescale for this process can be less than the age of the Solar
System, for plausible tidal parameters and moon masses.  However, as
we show numerically and analytically, the effects of the 2:1 and
3:1 resonant forcing terms between Nix and Charon complicate this
picture.  In the presence of Charon's tidal damping, the 2:1 term
forces Nix to migrate outward and the 3:1 term changes the
eccentricity damping rate, sometimes leading to eccentricity growth. We
conclude that this mechanism probably does not explain Nix and Hydra's
current orbits.  Instead, we suggest that they were formed in-situ with
low eccentricities.

We also show that an upper limit on Nix's migration speed sets a
lower limit on Pluto-Charon's tidal circularization timescale of 
$ >10^5 \,$ yrs. Moreover, Hydra's observed proper eccentricity may be explained
by the 3:2 forcing by Nix.

\end{abstract}

\section{Introduction}
\label{sec:int}

\cite{Weaver} discovered two small moons orbiting Pluto.
These moons---Nix and Hydra---are much less massive than Pluto's
major moon Charon, whose mass relative to Pluto's is
   $M_C/M_P
\simeq 0.12$ \citep{BGYYS06}, whereas Nix and Hydra's masses 
are
$M_N/M_P=(4\pm 4)\times 10^{-5}$ and
$M_H/M_P=(2\pm 4)\times 10^{-5}$, based
on a 4-body fit to the observed positions \citep{TBGE07}.\footnote{
Although these masses are consistent with zero, the brightness of the moons
implies that $M_N,M_H\sim 1-50\times 10^{-5}M_P$ 
for reasonable densities and albedoes \citep{Weaver}.}
 \cite{Weaver} also found that
(a) the orbits of Nix and Hydra  are nearly circular and coplanar with
Charon; and (b) the period ratios of Charon:Nix:Hydra are nearly
1:4:6.  By analyzing earlier data taken over a year-long interval and
fitting the data to Keplerian orbits around a point mass,
\cite{BGYYS06} showed that (a) Nix and Charon's eccentricities are
consistent with zero, with $e_N=0.0023(21)$ and $e_C=0.00000(7)$ where
brackets denote 1-$\sigma$ errors in the trailing digits, but Hydra's
formally is not, with $e_H=0.0052(11)$; and (b) the period ratio of
Charon:Nix:Hydra is 1:3.8915(2):5.9817(2).
Recently \cite{TBGE07} performed a full four-body fit to the moons' positions,
and thereby obtained slightly different orbital parameters.
We discuss some of their results in \S \ref{sec:discussion}.

Hopefully, Nix and Hydra's remarkable orbital properties can teach us something
about how they formed, perhaps shedding light on the
formation of Kuiper belt objects in general.  The near circularity and
coplanarity is surprising.  If Nix and Hydra formed in the collision
that produced Charon, they likely had high eccentricities and
inclinations just after formation.  Nix and Hydra cannot damp their
eccentricities through tidal interactions with Pluto---the
corresponding damping times, 
{for typical parameters, are at least $10^3$ times longer than
the age of the Solar System \citep[also see][]{Stern}.\footnote{This
is assuming monolithic bodies. For Nix, the e-folding time can be
brought to within a few Gyrs if it is a strengthless rubble pile and
has both tidal Love number and tidal $Q$ factor  of order unity.}}

\cite{WardCanup06} propose a scenario that not only accounts for Nix
and Hydra's low eccentricities and inclinations, but also accounts for
their near-resonant orbits.  In their ``forced resonant migration''
scenario, Nix and Hydra formed in the impact that produced Charon.
Immediately after impact, all three moons were much closer to Pluto
than they are today, and Nix and Hydra's eccentricities were damped by
a disk of post-impact debris.  Subsequently, Charon migrated outwards
by raising tides on Pluto, and it forced Nix and Hydra to migrate as
well because they were trapped in its 4:1 and 6:1 corotation
resonances.  Just before Charon stopped migrating, it left Nix and
Hydra at their current positions.  Although this is a neat scenario,
we show in a forthcoming paper (Lithwich \& Wu, in preparation) that
it cannot work.  The difficulty is that Charon cannot simultaneously
force both Hydra and Nix to migrate.  To transport Nix, Charon's
eccentricity must satisfy $e_C<0.024$; otherwise, the 4:1 corotation
resonance is destroyed by resonance overlap, as we prove with
numerical simulations.  But to transport Hydra, it is required that
$e_C> 0.1$; otherwise, the width of the 6:1 resonance is so small that
the time for the resonance to sweep across Hydra is shorter than the
libration time within the resonance.  Because these two constraints
are incompatible, the \cite{WardCanup06} scenario cannot work.

But Nix and Hydra can damp their eccentricities in a different way
that has not been considered previously: they can transfer their
eccentricity to Charon via secular interactions, and since Charon
damps its own eccentricity relatively quickly through its own tidal
interactions with Pluto, this process can damp eccentricities faster
than the direct interaction of Nix or Hydra with Pluto.  We shall show
below that the timescale for this process for Nix is
(eq. [\ref{eq:gm}])
\be
 2.7 \tau_C {M_C\over M_N},
\label{eq:tauint}
\ee
where $\tau_C$ is Charon's eccentricity damping time.
 The value of $\tau_C$ is somewhat uncertain.  If one assumes
that Pluto and Charon are consolidated ice spheres with material
strength of $4\times 10^{10}$erg/cm$^3$, then $\tau_C\sim 5 (Q/100)$ Myr
\citep{G63,Dob97}, where $Q$ is either Pluto's or Charon's tidal
damping parameter. (Both Pluto and Charon contribute
comparably to $\tau_C$.)  Taking $\tau_C = 5$ Myr, equation
(\ref{eq:tauint}) implies that the damping timescale is less than the
age of the Solar System if $ M_N/M_P\gtrsim 3\times
10^{-4}$.  Given the uncertainties in $\tau_C$---for example,
 $Q$ could be $\ll 100$, or Charon might be a rubble pile instead
 of  a consolidated
 ice sphere, which would increase its tidal Love number and
 decrease $\tau_C$---it is quite possible that the timescale of 
 equation (\ref{eq:tauint}) is less than the age of the Solar System.

One is then tempted to conclude that Nix and Hydra could have formed
with relatively large eccentricities, and that these were damped away
as described above.  However, we shall show below that, for Nix, the
2:1 and 3:1 resonant forcing terms with Charon severely complicate
this picture. 

\section{Pluto, Charon and Nix}

\subsection{N-body Simulation}
\label{sec:nbody}

\begin{figure}
\hspace{-1cm}\vspace{-2cm}\includegraphics[width=.59\textwidth]{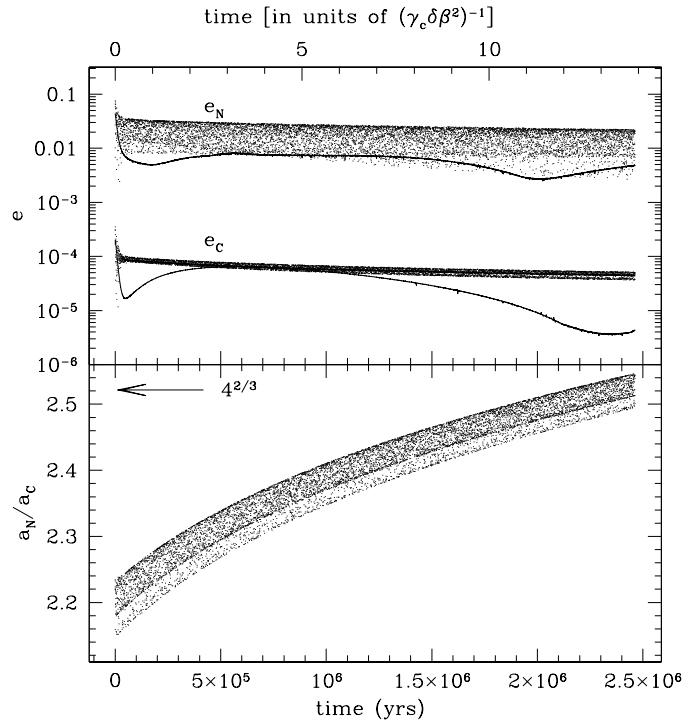}
\caption{N-body simulation of Pluto, Charon, and Nix with tidal
damping on Pluto and Charon. The simulation had tidal
circularization time $\tau_C=100$yrs;
for more realistic values of $\tau_C$,
the bottom time axis should be rescaled by $\tau_C/100$yrs.
  The top time axis is in units of the
damping time of the slowly damped secular mode (\S \ref{sec:damp}).
  In
the top panel, the points depict Charon and Nix's instantaneous 
eccentricities in Jacobi coordinates, and the solid lines depict their proper
 eccentricities.
 The lower panel displays the semi-major
axis ratio between Nix and Charon, with the arrow indicating the
nominal position of the $4:1$ resonance.  } \label{fig:nbody}
\end{figure}

Figure \ref{fig:nbody} shows the result of an
 N-body simulation of Pluto, Charon,
and Nix, with tidal damping acting on Pluto and Charon.  The N-body
integration is performed with the SWIFT package \citep{LD94}, using
the hierarchical Jacobi symplectic integrator of \cite{Beust}, and
supplemented with an algorithm for tidal damping.
 Nix and Charon have masses
$M_N=1.5\times 10^{-4}M_P$ and $M_C=0.1M_P$, respectively.
To speed up the simulation, we set the circularization time of 
the Pluto-Charon binary to $\tau_C=100$ yrs.  Although this
is shorter than the true $\tau_C$ by many orders of magnitude,
we shall show that all relevant timescales scale with $\tau_C$.
Hence for more realistic values of $\tau_C$, one need only
rescale the bottom time axis by the factor $\tau_C/100$ yrs.

 The 
points in the top panel show Charon and Nix's instantaneous eccentricities
in Jacobi coordinates,
 and the lines show their proper eccentricities.
To be more precise, the lines are  the time-average
of one component of the eccentricity vectors, averaged over 100 days, which
 is a crude way to remove the short-term noise from
the eccentricities.
We start Charon on a circular orbit at its current location (with a period
of 6.4 days),
and Nix on an orbit with an instantaneous eccentricity $0.05$ and
with $a_N/a_C \sim 2.2$, where $a$ is the semi-major axis.
At early times, Nix's proper eccentricity decays.  The top time
axis is in units of the damping time of the slowly damped
secular mode 
(eq. [\ref{eq:gm}], which is equivalent to eq. [\ref{eq:tauint}]).
The figure shows that Nix's eccentricity initially damps on a timescale much faster
than equation (\ref{eq:tauint}).  But at later times, Nix's eccentricity evolves
in a complicated manner:
it increases for a while before decaying again.

The bottom panel of Figure \ref{fig:nbody} shows the ratio of semimajor axes.
Charon's semimajor axis $a_C$ (not shown) remains virtually constant.
Nix's semimajor axis slowly increases throughout the simulation,
and crosses through the nominal position of the 4:1 resonance
(marked by the arrow).

\subsection{Simplified Model}

\label{subsec:not}

In the remainder of this section, we explain the complicated
behavior shown in Figure \ref{fig:nbody}.  
In this subsection we introduce a simplified model, 
keeping only a small number of terms in the disturbing function,
and show that a numerical integration of the simplified model
qualitatively reproduces the behavior of Figure \ref{fig:nbody}.
In \S\S \ref{sec:damp}-\ref{sec:2:1},
we isolate the individual terms within the simplified model,
and explain their effects.  The reader uninterested in the technical
details may skip to a summary of our findings in \S \ref{sec:sum}.

We adopt Jacobi orbital elements, in which Nix's elements are relative
to the center-of-mass of the Pluto-Charon binary, and Charon's
elements are relative to Pluto:
\be
\{a_j,\lambda_j,e_j,\pomega_j\} \ , \ \ j\in \{C,N\} \ 
\ee
where subscript $C$ is for Charon, or to be more accurate the
Pluto-Charon binary, and $N$ is for Nix.  Standard treatments of the
planetary equations adopt heliocentric (equivalent here to
Plutocentric) elements \citep{MD99}.  But the present problem is more
suited to Jacobi elements.  We show in the Appendix that using Jacobi
elements changes the standard treatment in two ways.  First, instead
of the usual indirect term in the disturbing function, there is a
different correction to the direct term, given by equation
(\ref{eq:df}) to first order in the ratio of Charon's to Pluto's mass
($M_C/M_P\sim 0.1$).  And second, the appropriate masses must be used;
for example, it is the reduced mass of the Pluto-Charon binary that
enters into equation of motion.  We ignore this second change in the
body of the paper because we only seek an accuracy of $\sim
M_C/M_P\sim 10\%.$\footnote{Adopting Plutocentric elements would give
corrections relative to the Jacobi elements that are significantly
larger than $M_C/M_P$.  To appreciate this, consider what would happen
if the Pluto-Charon binary was very tight.  Then Pluto's large reflex
velocity relative to the binary's center-of-mass would enter into
Nix's Plutocentric elements, even though Nix would be weakly perturbed
by the non-pointlike nature of the binary.  Using Jacobi elements
avoids this effect.  }

{We model the gravitational forces on Nix and Charon with the
truncated Hamiltonian
\be
H=H_{{\rm unp},C}+H_{{\rm unp},N}
+H_{{\rm sec},C N}+H_{2:1,C N}+H_{3:1,CN} \ .
\label{eq:bigham}
\ee
These terms and their associated coefficients are defined in Tables
\ref{tab:ham} and \ref{tab:dist}, respectively, 
where $M_j$ are the masses, $\mu_C\equiv M_C/(M_C+M_P)$,
\be
z_j\equiv e_j e^{i\pomega_j}
\ee
are the complex eccentricities, and we choose units
 so that
\be
\gn (M_P+M_C)=1 \ .
\ee

The equations of motion are given by Hamilton's equations\footnote{
The exact equations are given by equations
(\ref{eq:adot})-(\ref{eq:lamdot}) and, in place of equation
(\ref{eq:ham}), $d\zeta_j/dt=-i\partial H/\partial \zeta_j^*$, where
$\zeta_j\equiv (M_j\sqrt{a_j})^{1/2}
(1-\sqrt{1-e_j^2})^{1/2}e^{i\pomega_j}$ is a complex canonical
variable \citep{Ogilvie07}.  In converting to equation (\ref{eq:ham}),
we set $\zeta_j=(M_j\sqrt{a_j}/2)^{1/2}z_j$, valid to leading order in
$e_j$, and assume that $a_j$ is constant, because corrections caused
by varying $a_j$ are higher order in eccentricity.  \label{foot:canon}
} \beqn {dz_j\over dt}&=& - {2i\over M_j\sqrt{a_j}}{\partial H\over
\partial z_j^*}
\label{eq:ham} 
\\
{da_j\over dt}&=&-{2\sqrt{a_j}\over M_j}{\partial H\over \partial
\lambda_j}
\label{eq:adot}
\\
{d\lambda_j\over dt} &=& {2\sqrt{a_j}\over M_j}{\partial H\over\partial a_j}
\label{eq:lamdot}
%
 \ , \ \ \ j\in\{C,N\} \
\eeqn
For $d\lambda_j/dt$, it suffices in this paper to consider only the
contribution of the unperturbed energies $H_{{\rm unp},j}$, i.e.
\be
{d\lambda_j\over dt} =n_j \ , \label{eq:lamdot2}
\ee
where
\be
n_j\equiv a_j^{-3/2}
\ee
is the mean motion.

Both the 2:1 and the 3:1 resonant forcing terms in the Hamiltonian
play important roles even when Nix and Charon are not particularly
close to the nominal locations of those resonances.  

We model the tidal damping of Charon's eccentricity by adding the term
\be
{dz_C\over dt}\Big\vert_{\rm tide}\equiv -\gamma_C z_C
\label{eq:tide}
\ee
to the equation for $z_C$, where 
\be
\gamma_C\equiv 1/\tau_C
\ee
is Charon's eccentricity damping rate in the absence of Nix.

\begin{table*}[t]
\centering
\begin{minipage}{160mm}
 \caption{ Terms in Charon-Nix Hamiltonian Used in This
 Paper\tablenotemark{a}}
\begin{tabular}{l | l}
\hline
$H_{{\rm unp},C}=-{M_C\over 2a_C}$& Charon's unperturbed energy
\\
$H_{{\rm unp},N}=-{M_N\over 2a_N}$& Nix's unperturbed energy
\\
$H_{{\rm sec},C N}= -\mu_C{ M_N\over a_N}\left(
g_1\left(|z_C|^2+|z_N|^2\right) +{1\over 2}g_2\left(z_C z_N^*+z_C^*z_N
\right)
\right)$
 & 
secular interaction energy
 \\  
$H_{2:1,CN}=-\mu_C{M_N\over   2 a_N}
\left(
g_3z_C^*+g_4z_N^*
\right)e^{i(2\lambda_N-\lambda_C)}
+c.c.$ & 2:1 resonant interaction energy
\\
$H_{3:1,CN}=-\mu_C{M_N\over 2a_N}\left( g_5z_C^{*2} +
g_6z_C^*z_N^*+g_7z_N^{*2}\right) e^{i(3\lambda_N-\lambda_C)}+c.c. $ &
3:1 resonant interaction energy
\end{tabular}
\label{tab:ham}
\tablenotetext{a}{
``$c.c.$'' denotes complex conjugate of preceeding term.
Coefficients $g_i$ are listed in Table
\ref{tab:dist}.  }
\end{minipage}
\end{table*}

\begin{table*}[t]
\centering
\begin{minipage}{160mm}
 \caption{ Coefficients in Charon-Nix Hamiltonian\tablenotemark{a}}
\begin{tabular}{l l}
\hline
$g_1=\alpha b_{3/2}^{1}/8$&$=0.0815$
\\
$g_2=-\alpha b_{3/2}^{2}/4$&$= -0.0792$
\\
$g_3=(-2-\alpha D/2)b_{1/2}^2$&$=-0.390$
\\
$g_4 =(3/2+\alpha D/2)b_{1/2}^1-2\alpha$&$=0.0811$
\\
$g_5=(21/8+5\alpha D/4+\alpha^2D^2/8)b_{1/2}^3$&$=0.315$
\\
$g_6=(-5-5\alpha D/2-\alpha^2D^2/4)b_{1/2}^2$&$=-1.41$
\\
$g_7=(17/8+5\alpha D/4+\alpha^2D^2/8)b_{1/2}^1-27\alpha/8$&$=0.186$
\end{tabular}
\label{tab:dist}
\tablenotetext{a}{
The coefficients are given in Appendix B of \cite{MD99}, except that
we include the effective indirect term for Jacobi coordinates
(eq. [\ref{eq:df}]). The Laplace coefficients $b_s^j$ are functions of
$\alpha=a_C/a_N$, and $D\equiv d/d\alpha$. In the numerical
expressions, we set $\alpha=4^{-2/3}$.  }
\end{minipage}
\end{table*}

Figure \ref{fig:ham} shows the result from numerically integrating
these equations of motion, {where the $g_i$ coefficients in the 
Hamiltonian  are evaluated at the 4:1 location (Table
\ref{tab:dist}).} 
The evolution in this simplified model is qualitatively similar to
that seen in the N-body integration of Figure \ref{fig:nbody}.  The
 eccentricities initially decay, then rise, and finally decay
again. And  Nix's semimajor axis increases with time.

\begin{figure}
\hspace{-.8cm}\vspace{-2.6cm}\includegraphics[width=.57\textwidth]{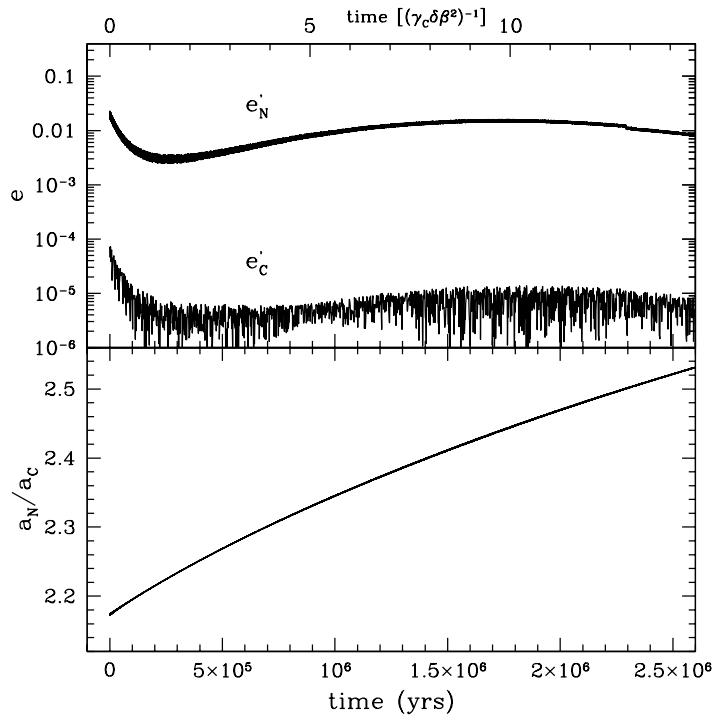}
\caption{Numerical integration of the simplified model.
The curves show the result from integrating equations (\ref{eq:ham}),
(\ref{eq:adot}), and (\ref{eq:lamdot2}) with tidal damping
(eq. [\ref{eq:tide}]), and with the Hamiltonian given by eq.
(\ref{eq:bigham}).  The masses of Pluto, Charon, and Nix and the tidal damping
rate of Charon are all the same as in the N-body simulation of Figure
\ref{fig:nbody}.  
The top panel shows the eccentricities after subtracting off the
complex eccentricities forced by the 2:1 resonance
(eqs. [\ref{eq:zc21}]-[\ref{eq:zn21}]).  The evolution seen in this
figure is qualitatively similar to that seen in Figure
\ref{fig:nbody}: the eccentricities first decay, then rise, then decay
again.  And Nix is pushed out by Charon.  }
\label{fig:ham}
\end{figure}

\subsection{Secular Evolution With Tidal Damping}

\label{sec:damp}

To explain the behavior of the simplified model, 
we consider first the effect of only the secular term
($H=H_{{\rm sec},CN}$), together with tidal damping
(eq. [\ref{eq:tide}]).
The equations of motion are then
\beqn
{d\over dt}
\left(\begin{array}{c}z_C \\z_N\end{array}\right)
= i\omn
\left(\begin{array}{cc}\delta  & -\beta\delta \\ -\beta & 1\end{array}\right)
\left(\begin{array}{c}z_\C \\z_N\end{array}\right)
-\gamma_C\left(\begin{array}{c}z_\C \\ 0\end{array}\right)
\label{eq:zdot}
\eeqn
where
\beqn
\delta \equiv {M_N\over M_C}{\sqrt{a_N}\over \sqrt{a_C}}  \label{eq:delta}  , \ \ 
\omn  \equiv  2n_N {\mu_\C}g_1 \label{eq:omn} , \ \ 
 \beta \equiv  -{g_2\over  2 g_1} \label{eq:beta}  \ .
\eeqn
 Equation (\ref{eq:zdot}) is a linear
equation with constant coefficients.  It is a simple exercise in
linear algebra to solve it \citep{WG02}.  Since $\gamma_C$ is much
smaller than any of the frequencies in the problem, we first consider
what happens in the absence of damping by setting $\gamma_C=0$.
Setting $(z_\C,z_N)\propto e^{i\omega t}$, the eigenvalues and
eigenfunctions are \beqn \omega_\pm&=&{\omn \over 2}\left( \delta+1\pm
(\delta-1)\sqrt{1+\epsilon} \right) \label{eq:efn} \\ %
\left(\begin{array}{c}z_\C \\ z_N \end{array}\right)_\pm &\propto&
\left(\begin{array}{c}1\pm\sqrt{1+\epsilon} \\
\sqrt{\epsilon/\delta}\end{array}\right) \label{eq:w1w2} \eeqn where
\beqn \epsilon &\equiv& {4\delta\beta^2/ (1-\delta)^2} \eeqn With
non-zero $\gamma_C$, both the $+$ and $-$ modes are damped.  The
damping rates are found by calculating the imaginary part of the
eigenvalues of the full equation (eq. [\ref{eq:zdot}]), and expanding
to first order in $\gamma_C$, which yields the damping rates
\beqn
\gamma_{\pm}=
{\gamma_{C}\over 2}\left( 1\pm
\left(1+{\epsilon}\right)^{-1/2} 
 \right) \ . \label{eq:gpm}
\eeqn

The physical meaning of the above results becomes clearer in the limit
that $\delta\ll 1$.  For the eigenvalues and eigenfunctions in the
absence of damping, consider first what happens when Charon's
eccentricity is held fixed, $z_C=$constant$\equiv \hat{z}_C$. Then
equation (\ref{eq:zdot}) shows that $z_N=\beta\hat{z}_C+z_{N,fr}$,
where $\beta\hat{z}_C$ is the constant forced eccentricity, and the
free eccentricity $z_{N,fr}$ has constant amplitude and a phase that
precesses at frequency $\omn $.  Similarly, if one artificially held
Nix's eccentricity fixed, $z_N=$constant$\equiv \hat{z}_N$, then
Charon would have forced eccentricity $=\beta\hat{z}_N$, and its free
eccentricity would precess at frequency $\omn\delta $.  When both
$z_C$ and $z_N$ are allowed to evolve freely, there are two modes.  In
one of them, which can be called the CfN mode (``Charon forces Nix''),
Nix's eccentricity is equal to its value forced by Charon, $z_N=\beta
z_C$. From Charon's equation of motion, we see that this mode has
frequency $\omn\delta(1-\beta^2)$, comparable to Charon's free
precession frequency $\omn\delta$; therefore the CfN mode has
\beqn
\omega_+&=&\omn \delta(1-\beta^2) \\
\left(\begin{array}{c}z_C \\z_N\end{array}\right)_+&\propto&
\left(\begin{array}{c}1 \\\beta\end{array}\right) \label{eq:evp} \ ,
\eeqn
in agreement with the small $\delta$ limit of the $+$ mode in
equations (\ref{eq:efn})-(\ref{eq:w1w2}).

In the other mode (NfC =``Nix forces Charon''), Nix has a free
eccentricity, so it is freely precessing.  Nix's eccentricity tends to
drive $z_C$ to precess around its forced value $\beta z_N$.  But since
Nix precesses much faster than Charon---by the factor $1/\delta\gg
1$---Charon can only reach an eccentricity of $|z_C| \sim \beta\delta
z_N$ in the time that Nix undergoes a full precession period.  Since
$|z_C|\ll |z_N|$, Nix's equation of motion is $dz_N/dt\approx i\eta
z_N$, showing that this mode has frequency equal to Nix's free
precession frequency $\omn$.  And Charon's equation of motion shows
that $z_C=-\beta \delta z_N$, so the NfC mode has
\beqn
\omega_-&=&\omn  \label{eq:omm}\\
\left(\begin{array}{c}z_C \\z_N\end{array}\right)_-&\propto&
\left(\begin{array}{c}-\beta\delta \\ 1\end{array}\right) \label{eq:evm}
\eeqn
in agreement again with equations (\ref{eq:efn})-(\ref{eq:w1w2}) for
$\delta\ll 1$.

When tidal damping is active, the damping rate of the CfN mode in the
$\delta\ll 1$ limit is just that of Charon in isolation
\be
\gamma_+ = \gamma_C \ .
\ee
For the NfC mode, we take advantage of the fact that $|z_C|\ll|z_N|$.
Nix's approximate equation $dz_N/dt\approx i\eta z_N$ implies
$z_N=\hat{z}_Ne^{i\eta t}$, where $\hat{z}_N$ is (nearly)
constant. Charon's equation is then $dz_C/dt\approx
-i\eta\beta\delta\hat{z}_Ne^{i\eta t}-\gamma_Cz_C$; its solution after
a few damping times of the CfN mode ($t\gg 1/\gamma_C$) is
$z_C=-\beta\delta(1+i\gamma_C/\eta)\hat{z}_Ne^{i\eta t}$.  We now
consider the full equation for $z_N$, and substitute into this
equation both the above $z_C$ and the zeroth order solution
$z_N=\hat{z}_Ne^{i\eta t}$, where $\hat{z}_N$ is now allowed to be
time-dependent, yielding
$d\hat{z}_N/dt=(i\eta\beta^2\delta-\gamma_C\delta\beta^2)\hat{z}_N$.
Therefore $|z_N|$ damps at the rate
\be
\gamma_-=\gamma_C \delta\beta^2 \ . \label{eq:gm} \ ,
\ee
in agreement with the small $\delta$ limit of equation (\ref{eq:gpm}).
This gives the damping time quoted in the introduction
(eq. [\ref{eq:tauint}]).

To summarize, the purely secular evolution has two normal modes, one
of which (CfN) decays very quickly.  The second mode (NfC) is much
more slowly damped.  As shown in the introduction, the timescale to
damp this mode can be shorter than the age of the Solar System for
reasonable values of $\gamma_C$ and $M_N/M_C$.

\subsection{Adding in the 3:1 Resonant Forcing Term}

The purely secular evolution changes markedly when the 3:1 forcing
term $H_{3:1,CN}$ is included.  In Appendix B, we show that including
the 3:1 term between Charon and Nix alters the damping rate of the
slowly damped NfC mode from $\gamma_-$ (eq. [\ref{eq:gm}]) to
$\gamma_{3:1}$ (eq. [\ref{eq:gamma31}]).  Figure \ref{fig:3to1} plots
the ratio $\gamma_{3:1}/\gamma_-$ as a function of the semimajor axis
ratio.  Remarkably, the 3:1 term has an effect at large distances from
the location of the nominal 3:1 resonance, $(a_N/a_C)^{3/2}=3$.

Far beyond the nominal 4:1 resonance, $(a_N/a_C)^{3/2} \gg 4$, the 3:1
resonant forcing term has little effect
($\gamma_{3:1}/\gamma_-\rightarrow 1$).  But at Nix's current location
$(a_N/a_C \simeq 4^{2/3})$, the damping rate is reduced to only $7\%$
of the purely secular rate. Furthermore, in the range
$3.4<(a_N/a_C)^{3/2}<3.9$ the damping rate is negative: instead of
damping there is exponential growth.  For still smaller values of
$a_N/a_C$ the damping rate is very large and positive. {These
behaviours are also confirmed by the full N-body integration
(Fig. \ref{fig:nbody}).}

\begin{figure}
\hspace{-.5cm}\vspace{-2cm}\includegraphics[width=.55\textwidth]{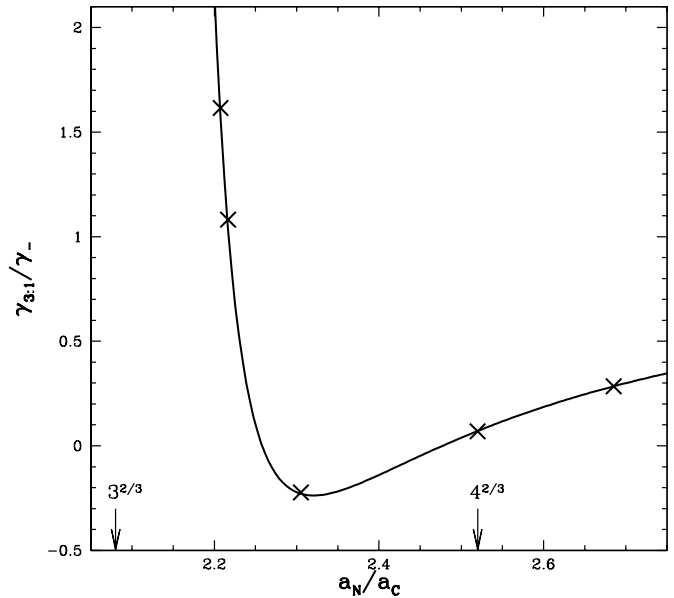}
\caption{{Effect of the 3:1 resonance on the eccentricity damping rate:}
The curve shows the ratio of $\gamma_{3:1}$ (the secular damping rate
including the 3:1 resonance) to $\gamma_-$ (the rate without the 3:1
resonance), where $\gamma_{3:1}$ is given by equation
(\ref{eq:gamma31}) and $\gamma_-$ by equation (\ref{eq:gm}).  In the
range $2.3<a_N/a_C<2.5$ (i.e., $3.4<(a_N/a_C)^{3/2}<3.9$) the damping
rate is negative, implying exponential growth.  The x's show
results from numerical integrations of equation (\ref{eq:3to1eq}),
showing good agreement with the analytic expression. }
\label{fig:3to1}
\end{figure}

\subsection{Effect of the 2:1 Resonant Forcing Term}
\label{sec:2:1}

\begin{figure}
\centering
\hspace{-.7cm}\vspace{-2cm}\includegraphics[width=.52\textwidth]{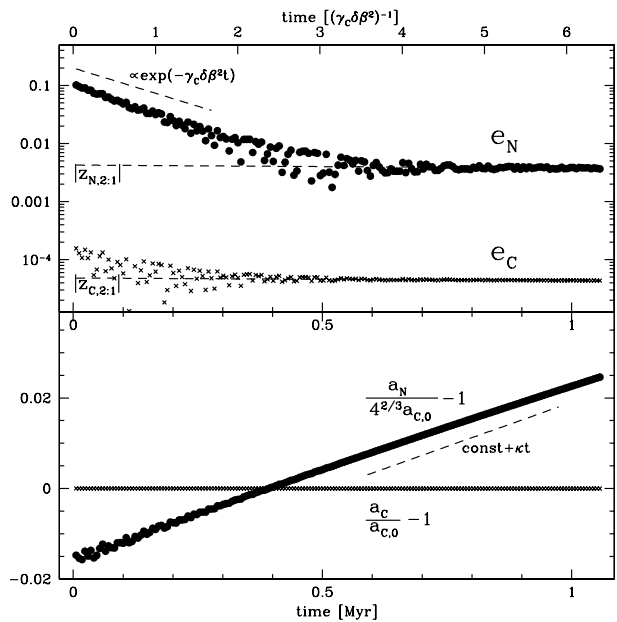}
\caption{Numerical simulation of Charon and Nix, in which Charon and 
Nix are coupled both secularly and via the 2:1 resonance, and tides
damp Charon's eccentricity.  Top panel shows eccentricities
($e_N\equiv |z_N|, e_C\equiv |z_C|$) and bottom panel shows semimajor
axes relative to Charon's initial semimajor axis $a_{C,0}$, confirming
the solution given in equations (\ref{eq:zcznsol}) and
(\ref{eq:acansol}).  In the top panel, the slowly damped mode damps at
the rate $\gamma_C\delta\beta^2$, and after 2-3 damping times the
eccentricities reach their { resonantly} forced values; the
dashed lines labelled $|z_{N,2:1}|,|z_{C,2:1}|$ are the forced
eccentricities given by equations (\ref{eq:zc21})-(\ref{eq:zn21}).
The dashed line in the bottom panel shows that Nix's semimajor axis
increases at the rate $\kappa$ (eq. [\ref{eq:kap}]). {The
addition of the 3:1 resonant forcing term will qualitatively change
the eccentricity evolution, but not the semi-major axis evolution.} }
\label{fig:nohydra}
\end{figure}

In this subsection, we solve the simplified model of \S
\ref{subsec:not} when Charon and Nix are coupled both secularly and
via the 2:1 resonance, and tides damp Charon's eccentricity. We
discard the 3:1 resonant forcing term.  The equations for $\lambda_C$
and $\lambda_N$ are given by equation (\ref{eq:lamdot2}), the
eccentricity equations are given by equation (\ref{eq:zdot}) with the
following term added to the right-hand side,
\beqn
{d\over dt}
\left(\begin{array}{c}z_C \\z_N\end{array}\right)\Big\vert_{2:1}
\equiv 
i\mu_Cn_N\left(\begin{array}{c} g_3 \delta \\ g_4\end{array}\right)
e^{i(2\lambda_N-\lambda_C)} \ ,
\label{eq:zdot21}
\eeqn
and the semimajor axis equations are
\beqn
{d\over dt}
\left(\begin{array}{c}a_C \\a_N\end{array}\right)
=
\nonumber
\\
i\mu_Cn_N 
\left(\begin{array}{c}-\delta a_C \\2a_N\end{array}\right)
(g_3z_C^*+g_4z_N^*)
e^{i(2\lambda_N-\lambda_C)}+ c.c.
\label{eq:adot21}
\eeqn

Since $a_C,a_N$ vary by only a small amount ($O(\mu_C e_N)$ or
smaller) on the 2:1 timescale, we may treat $n_C,n_N$ as constants in the equations for the
mean longitudes, which are then trivially solved
\beqn
\lambda_C&=&n_Ct+{\rm const} \\
\lambda_N&=&n_Nt+{\rm const} \ .
\eeqn

The general solution of the eccentricity equation is the sum of the
two homogeneous (or ``free'') solutions given in \S \ref{sec:damp} and
the particular (or ``forced'') solution that is driven at frequency
\be
n_{2:1}\equiv 2n_N-n_C \ .
\ee
 Discarding
the rapidly damped free solution (CfN), leaves
\be
\left(\begin{array}{c}z_C \\z_N\end{array}\right)
=z_{N,fr}(t)\left(\begin{array}{c}-\beta\delta  \\ 1\end{array}\right)+
\left(\begin{array}{c}z_C \\z_N\end{array}\right)_{2:1} 
\label{eq:zcznsol}
\ee
where $z_{N,fr}(t)$=const$\times e^{i(\eta-\gamma_C\delta\beta^2)t}$
and
\beqn
z_{C,2:1}&=&
\mu_C
g_3\delta
{n_N\over n_{2:1}}
 e^{in_{2:1} t}
\left(1+i{\gamma_C\over n_{2:1}}\right)
\label{eq:zc21}
\\
z_{N,2:1}&=&\mu_Cg_4{n_N\over n_{2:1}}e^{in_{2:1}t}
\label{eq:zn21}
\eeqn
to first order in $\gamma_C$. Even though the $O(\gamma_C)$ correction
to $z_{C,2:1}$ is very small in absolute value, it plays an important
role in the evolution of the semimajor axes because of its imaginary
coefficient.  Secular terms have been discarded from the forced
solution, which is appropriate because $\eta\ll |n_{2:1}|$.

We may now substitute these eccentricities into equation
(\ref{eq:adot21}).  The free eccentricities produce rapidly
oscillating terms $\propto e^{i(n_{2:1}+\eta)t}$ that lead to small
variations of $a_C,a_N$.  But the forced eccentricities have a more
interesting effect: they induce a slow secular change,
\beqn
{d\over dt}
\left(\begin{array}{c}a_C \\a_N\end{array}\right)
&=&
\kappa
\left(\begin{array}{c}-\delta a_C/2 \\ a_N\end{array}\right)
\label{eq:acansol}
\\
\kappa&\equiv&
4\delta\gamma_C
\left(
\mu_Cg_3{n_N\over n_{2:1}}
\right)^2 \ . 
\label{eq:kap}
\eeqn
In the time it takes Nix's free eccentricity to damp, its semimajor
axis increases by the factor $\kappa/(\beta^2\delta\gamma_C)=0.6
\mu_C^2\simeq 0.6\%$ at $a_N=4^{2/3}a_C$; $a_N$ continues to increase
even after the free eccentricities have damped away.

Figure \ref{fig:nohydra} shows a numerical integration of the
equations of motion given in this subsection, with the same parameters
for the N-body integration shown Figure \ref{fig:nbody}.  The evolution
seen in the Figure confirms the solutions given in equations
(\ref{eq:zcznsol}) and (\ref{eq:acansol}).

\subsection{Summary of Pluto, Charon, and Nix's Evolution}
\label{sec:sum}

We have explained the behavior seen in the N-body
simulation  of
Figure \ref{fig:nbody}.
There are three important  types of interactions between 
Charon and Nix: secular, 3:1 forcing, and 2:1 forcing.
Secular interactions lead to two damped normal modes.
One of these (CfN=Charon forces Nix) is rapidly damped away on the
 timescale of Charon's tidal damping time $\tau_C=1/\gamma_C$.
This is too short to be seen in Figure \ref{fig:nbody}.  The second normal
mode (NfC=Nix forces Charon) is damped on a much longer timescale, 
at the rate $\gamma_-=\gamma_C\delta\beta^2\sim \gamma_CM_N/M_C$.
In this mode, Charon's proper eccentricity is forced by Nix to 
$e_C\sim (M_N/M_C)e_N$.
But purely secular effects do not suffice to explain the evolution seen in Figure \ref{fig:nbody}.
The 2:1 forcing term causes Nix's semimajor axis to increase at the rate $\kappa\sim 0.006\gamma_-$.
The 3:1 forcing term changes the damping rate of the NfC mode.  At early times in
Figure \ref{fig:nbody}, the 3:1 term forces the damping rate to be much faster than $\gamma_-$.
At later times, as Nix's semimajor axis increases, it enters a region where the 3:1 term
causes the eccentricities to grow, rather than damp (see Fig. \ref{fig:3to1}).  At even
later times, the proper eccentricities decay again, though at a rate less than $\gamma_-$.

Although we have ignored Hydra in our discussion, we describe its dynamics in 
Appendix \ref{sec:hydra}.

\section{Discussion}
\label{sec:discussion}

\subsection{Current Orbits of the Moons}
\label{subsec:keplerian}

Can Nix's current orbit be explained as the end state
of the evolution seen in Figure \ref{fig:nbody}? To answer this
question, we first need a better understanding of the current 
orbits of the moons.

\subsubsection{Nix's Eccentricity}

\cite{BGYYS06} fit Nix's orbit to a Keplerian orbit and found 
$e_N=0.0023(21)$.  \cite{TBGE07} fit all the moons' orbits simultaneously to
the results of 4-body integrations, and found for the best-fit solution
that Nix's eccentricity varied in time, with $0\leq e_N\leq 0.0272$.

As seen in Figure \ref{fig:nbody}, the instantaneous values of Nix's
eccentricity (shown as points), can be much
larger than the time-averaged eccentricity vector (shown as a line, for
an  averaging time of
100 days).
A clearer view of Nix's orbital state at the end of that simulation
 may be seen by taking 
 a Fourier transform of the eccentricity vectors around this time
(bottom panel of Figure \ref{fig:fourier}).
The forest of
high-frequency peaks at $\omega\gtrsim 0.1$/day are forced eccentricities.
They are associated with the resonant terms in the disturbing function.
For example, Nix's peak at $\omega=|2n_N-n_C|$ is the 2:1 forced
eccentricity described above (eq. [\ref{eq:zn21}]).
The low-frequency peak at $\omega\sim 0.0035$/day  is Nix's
secular eccentricity (which may also be called its ``free'' or ``proper''
eccentricity).
Even though the forced eccentricities dominate the instantaneous
eccentricity, they are not relevant if one wishes
to use Nix's current orbital state to infer something about its past.
This is because for fixed semimajor axes and masses, the forced
eccentricities are fixed.  It is only the secular eccentricity vector that
represents a true degree of freedom.  
Tides act
to damp the secular eccentricity, but they have no effect on the
forced eccentricity.  At the time
depicted in Figure \ref{fig:fourier} the CfN mode has damped away,  and
only the NfC mode remains.  The frequency of the NfC mode is 
$\omega_-=\eta$ (eq. [\ref{eq:omm}]), or $\omega_-\sim 0.004$/day, 
in agreement with the secular peak seen in the figure, i.e., the low-frequency
peak is the remains of the NfC mode as it is being damped by tides.

\begin{figure}
\hspace{-.5cm}\vspace{-2.2cm}\includegraphics*[width=0.55\textwidth]{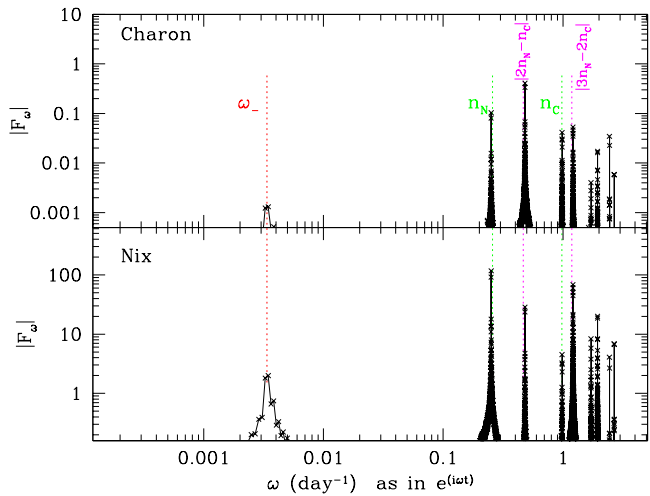}
\caption{Fourier decomposition of Charon \& Nix's eccentricity vector 
for the final state in Fig. \ref{fig:nbody}.  The
Fourier amplitude $F_\omega = \int (e \cos\pomega)\cos(\omega t) dt$
with orbital elements measured in Jacobi coordinates.
 The eccentricity of Charon is dominated by
the 2:1 forcing term, while Nix by a combination of short
term forcing. The peaks (indicated by $\omega_{-}$) are associated
with the slowly decaying secular mode (NfC) which precesses at $\omega_{-}$
(eq. [\ref{eq:evm}]). Their amplitude is a factor of $\sim 100$ below the total
eccentricity. }
\label{fig:fourier}
\end{figure}

In Figure \ref{fig:fit}, 
we  ``observe'' the orbital state  at the
end of the simulation in the manner of \cite{BGYYS06}
by fitting  Nix's orbit to a Keplerian ellipse
around a point mass at the barycenter with mass $M_P+M_C$ (where the value
of $M_P+M_C$ is to be found by the fit). We also fit Charon's orbit relative
to Pluto with a Keplerian ellipse.
We output the
positions of Charon and Nix as  functions of time from our numerical
integrator.   We then use the downhill simplex
method \citep{NR92} to search fits for the set of 5
parameters: $e\cos\pomega$, $e\sin\pomega$, $a$, $M_P+M_C$ and $\tau$
(epoch of periapse passage).
We find that the Keplerian-fit eccentricities are the secular parts of the total
eccentricity, with the instantaneous values being much greater but
averaged out in the fit.
Therefore, somewhat counterintuitively, the Keplerian fit of \cite{BGYYS06}
provides a more useful diagnostic of Nix's eccentricity than the 4-body
fit of \cite{TBGE07}.

\begin{figure}
\hspace{-.7cm}\vspace{-2cm}\includegraphics[width=0.55\textwidth]{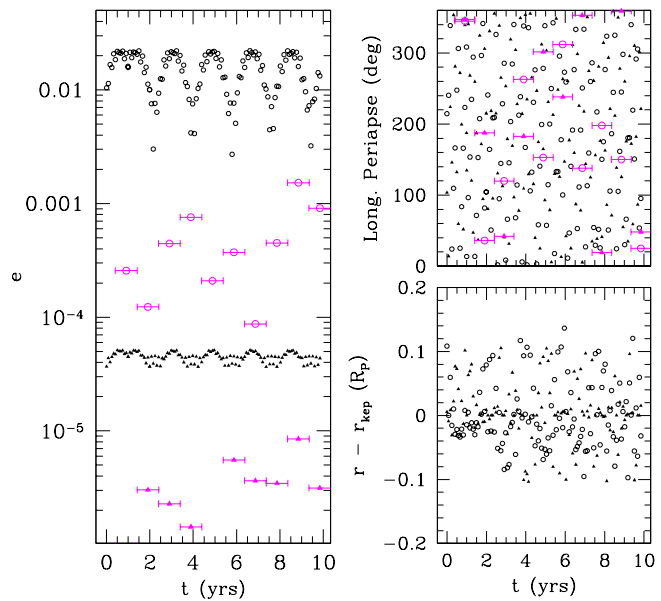}
\caption{Results of Charon \& Nix orbital fitting 
for the final state in
Fig. \ref{fig:nbody},
insisting on Keplerian  
ellipses . We output data (points for
Charon and open circles for Nix) once a month for 10 years, and do a
Keplerian fit every 12 months (solid triangles for Charon and open
triangles for Nix, error-bars in time-axis indicating span of
Keplerian fit). Eccentricities are shown in the left panel: $e_{\rm
Kep}$ fall well below the instantaneous values and are approximately
the secular values (obtained from Fig. \ref{fig:fourier}).  The
upper-right panel shows the value of $\pomega$, which varies on
orbital timescales and are not captured by the Keplerian fits.  The
lower-right panel displays the fitting residuals measured by $r-r_{\rm
kep}$ in unit of $R_P$. Charon's residuals are magnified by a factor
of $100$.}
\label{fig:fit}
\end{figure}

\subsubsection{Charon's eccentricity}

\cite{TBGE07} report that $e_C=0.00348(4)$
 when they do either a 4-body 
or a Keplerian fit, and that their earlier, much smaller
value for $e_C$ 
 \citep{BGYYS06} was incorrect.
We find this large value of $e_C$  very puzzling.  
Our theory predicts that Charon's eccentricity rapidly damps away by tides.
 More precisely, 
on the timescale $\tau_C\lesssim 10$ Myr, the CfN mode damps away.  After this
happens, Charon's secular eccentricity in the NfC mode is
 $e_C\sim (M_N/M_C) e_N$ as seen also in Figures \ref{fig:fourier}-\ref{fig:fit}.
 Thus Charon's eccentricity should be much smaller than the \cite{TBGE07}
 value.\footnote{Our theory predicts that Charon's secular eccentricity
 should be much too small to be observable.  But its forced
 eccentricity should be $\sim 0.3M_N/M_P\sim 10^{-5}$, forced
 primarily by the 2:1 resonance with Nix, and hence
 rapidly precessing at the 2:1 frequency (eq. [\ref{eq:zc21}], Figs. \ref{fig:fourier}-\ref{fig:fit}).
If this forced eccentricity is measured, one could infer from it Nix's mass.}   It is highly implausible that Charon's eccentricity was
 excited in the last $10$ Myr by some external event, such as flybys
 by passing Kuiper belt objects \citep{SBL03}.
In our view, the most plausible resolution of this puzzle is that the
high eccentricity of \cite{TBGE07} is incorrect; perhaps 
 inhomogeneities on 
the surface of Pluto or Charon are responsible for giving
an apparent eccentricity. Future observations should help to
resolve the puzzle.

\subsection{Constraints on Nix's Initial Orbit}
\label{subsec:nixonlyconstraint}

\begin{figure}
\centering
\hspace{-.9cm}\vspace{-2cm}\includegraphics[width=0.53\textwidth]{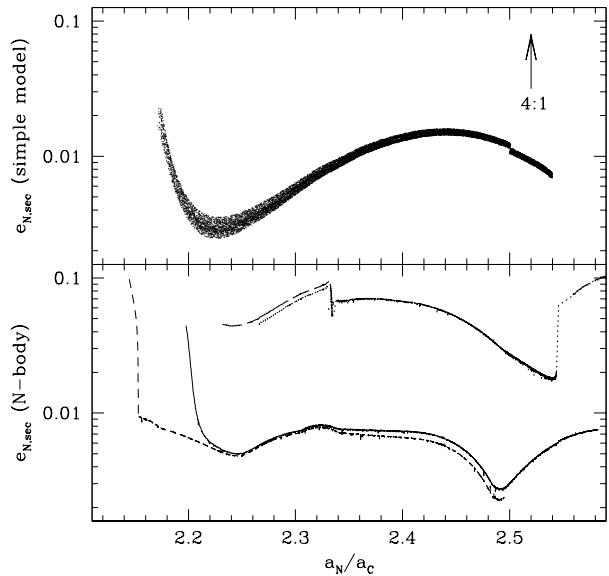}
\caption{Trajectory of Nix's orbit when tides damp Charon's eccentricity.
The top panel shows a numerical integration of the simplified model.  The
simulation is identical to the one shown in Figure \ref{fig:ham}, but here the
eccentricity is plotted versus semimajor axis ratio.
Similarly, the solid line in the bottom panel is the proper eccentricity 
from the N-body 
simulation shown in  Figure \ref{fig:nbody}, with the value of $a_N$ averaged
over 100 days.
The other curves  in the lower panel are from N-body simulations that started
with Nix both further in and further
out, but also with $e_N = 0.05$. The simplified model captures the
essential behavior of the system but differs in detail, 
partly because the Laplace coefficients are taken to be constants, 
and partly because our disturbing function is only accurate to $O(\mu_C)$.
 Nix's
current proper eccentricity is $\sim 0.002$
\citep{BGYYS06}. The nominal location of the 4:1 resonance
 is shown as an arrow. Trajectories
that have very low $e_{\rm N,sec}$ do not experience an appreciable
jump when crossing this resonance (confirmed by further integration
not shown here), while trajectories that have higher $e_{\rm sec}$
($\geq 0.02$) do.}
\label{fig:Charon_Nix_compareYoram}
\end{figure}

Let us consider the following scenario for explaining why Nix currently has
a very low proper eccentricity, $e_N\sim 0.002$ \citep{BGYYS06}:
perhaps Nix formed with a high eccentricity
 at $a_N\lesssim 2.25 a_C$, 
and then  tidal evolution  circularized its orbit
as it was migrated to its current location, as in Figure \ref{fig:nbody}.  Note that
Nix's proper eccentricity at the end of that simulation is close to the 
Keplerian-fit  value of  \cite{BGYYS06}.  
The bottom panel of Figure  \ref{fig:Charon_Nix_compareYoram} shows
two additional N-body  simulations with differing initial $a_N$.
It illustrates that if Nix started inward of $\sim 2.25a_C$, its proper eccentricity
is damped when it reaches its current location near the 4:1 resonance. But
if it started beyond $2.25a_C$, it would have retained much of its initial 
eccentricity.

The difficulty with this scenario is that 
to have migrated Nix from inward of $2.25 a_C$ to its current
position, we require that $\tau_C \leq 2\times 10^5 \, {\rm yrs}
(\mu_N/0.00015)$ (Fig. \ref{fig:nbody}), where $\mu_N\equiv M_N/M_P$.
Even for $\mu_N \sim 0.00015$, this requires a tidal damping $> 10$
times more efficient than the current estimate for the Pluto-Charon binary
\citep{Dob97}. 
This seems difficult unless Charon is a molten sphere with large Love
number $k_2$ (tidal distortion) and small $Q$ value. Moreover, it is a
strange coincidence that Nix is so close to the 4:1 location today,
since in our theory the outward migration of Nix does not pause at
4:1.

So unless the tidal dissipation time of Charon $\tau_C\lesssim
10^5$\, yrs, Nix will have to be initially deposited at its present semimajor axis
with
proper eccentricity $0.002$, its current measured value.
This places very stringent constraint for the formation scenario and
rules out all but formation in a disc. In the latter case, the
near 4:1 location is a result of migration in the disc (Shannon
et al., in preparation).

Conversely, the fact that Nix could not have migrated further than
from inward of the inner-most stable orbits ($a_N \sim 2.15 a_C$)
restricts $\tau_C$ to be $\geq 2\times 10^5$\, yrs.

\appendix
\section{Appendix A. Planetary Equations In Jacobi Coordinates}

The Hamiltonian for Pluto, Charon, and Nix is
\be
H={P_P^2\over 2 M_P}+{P_C^2\over 2 M_C} + {P_N^2\over 2 M_N} -{\gn
M_PM_C\over |\bld{R}_P-\bld{R}_C|} -{\gn M_NM_C\over
|\bld{R}_N-\bld{R}_C|} -{\gn M_NM_P\over |\bld{R}_N-\bld{R}_P|} \ ,
\ee
where $\bld{R}_j$ are position vectors from an arbitrary inertial
origin, $\bld{P}_j=M_jd\bld{R}_j/dt$ are the conjugate momenta, and
$M_j$ the masses.  In Jacobi coordinates, Charon's position is
measured relative to Pluto's, and Nix's position is measured relative
to the center-of-mass of the Charon-Pluto binary.  We transform to
Jacobi coordinates in two steps.  First, we employ the generating
function
\be
F=(\bld{R}_C-\bld{R}_P)\bld{\cdot}\bld{p}_{PC}+
 {M_C\bld{R}_C+M_P\bld{R}_P\over M_C+M_P} \bld{\cdot}\bld{P}_{PC}
\ee
to switch the coordinates of the Pluto-Charon binary from
$\bld{R}_P,\bld{R}_C,\bld{P}_P,\bld{P}_C$ to the binary's
center-of-mass and its relative position vector,
\beqn
\bld{R}_{PC}&\equiv& {M_C\bld{R}_C+M_P\bld{R}_P\over M_C+M_P} \\
\bld{r}_{PC}&\equiv& \bld{R}_C-\bld{R}_P \ ,
\eeqn
and their conjugate momenta  ($\bld{P}_{PC},\bld{p}_{PC}$),
yielding the new Hamiltonian
\beqn
H={P_{PC}^2\over 2 (M_P+M_C)}+{p_{PC}^2\over 2 M_{PC}} + {P_N^2\over 2
M_N} -{\gn (M_P+M_C)M_{PC}\over r_{PC}} -{\gn M_NM_C\over
|\bld{R}_N-\bld{R}_{PC}-(1-\mu_C)\bld{r}_{PC}|} -{\gn M_NM_P\over
|\bld{R}_N-\bld{R}_{PC}+\mu_C\bld{r}_{PC}|} \ ,
\eeqn
where 
\beqn
M_{PC}={M_PM_C\over M_P+M_C} 
\eeqn
is the reduced mass and
\beqn
\mu_C={M_C\over M_P+M_C} \ .
\eeqn
From Hamilton's equation for $d{\bld R}_{PC}/dt,d{\bld r}_{PC}/dt$, we
see that $\bld{P}_{PC}$ is the total momentum of the binary, and
$\bld{p}_{PC}$ is the momentum of the relative orbit.

We complete the transformation to Jacobi coordinates with the
generating function
\be
F=\bld{R}_{PC}\bld{\cdot}\bld{P}_{PCN}+({\bld{R}_N-\bld{R}_{PC}})\bld{\cdot}\bld{p}_N
\ee
which yields the new coordinate
\be
\bld{r}_N\equiv \bld{R}_N-\bld{R}_{PC} \ ,
\ee
as desired, as well as the new coordinate $\bld{R}_{PCN}\equiv
\bld{R}_{PC}$; the conjugate momenta are $\bld{p}_N\equiv\bld{P}_N$
and the total center-of-mass momentum
\be
\bld{P}_{PCN}\equiv \bld{P}_{PC}+\bld{P}_N \ .
\ee
Since this is constant ($\bld{R}_{PCN}$ does not appear in the Hamiltonian),
we may set it to zero.
The Hamiltonian in Jacobi coordinates is
\beqn
H(\bld{p}_{PC},\bld{r}_{PC};\bld{p}_N,\bld{r}_N)= {p_{PC}^2\over 2
M_{PC}}+{p_N^2\over 2M_N}\left(1+{M_N\over M_P+M_C} \right)
\\
-{\gn (M_P+M_C)M_{PC}\over r_{PC}}
-{\gn M_NM_C\over |\bld{r}_N-(1-\mu_C)\bld{r}_{PC}|} 
-{\gn M_NM_P\over |\bld{r}_N+\mu_C\bld{r}_{PC}|}
\ ,
\eeqn
Thus far, our treatment has been exact.  Henceforth, we drop the
factor $M_N/(M_P+M_C)$ in the second term above; although it is simple
to retain it, we drop it for notational convenience.  We choose units
so that
\be
G(M_P+M_C)= 1 \ .
\ee
By adding and subracting the term $-M_N/r_N$, the Hamiltonian may be
written as
\be
H(\bld{p}_{PC},\bld{r}_{PC};\bld{p}_N,\bld{r}_N)=H_{{\rm unp},PC}+H_{{\rm unp},N} + H_{\rm pert} \ ,
\ee
where
\beqn
H_{{\rm unp},PC}&=&{p_{PC}^2\over 2 M_{PC}}-{M_{PC}\over r_{PC}} 
\\
H_{{\rm unp},N}&=&{p_{N}^2\over 2 M_{N}}-{M_N\over r_{N}} 
\\
H_{\rm pert}&=& -{M_NM_C \over M_P+M_C}\left( {1 \over
|\bld{r}_N-(1-\mu_C)\bld{r}_{PC}|}-{1\over r_N} \right) -{M_NM_P \over
M_P+M_C}\left( {1 \over |\bld{r}_N+\mu_C\bld{r}_{PC}|}-{1\over r_N}
\right) \ , \label{eq:hpert}
\eeqn
Since we seek equations for the orbital elements, we transform
variables to $\{a_{PC},\lambda_{PC},e_{PC},\pomega_{PC}\}$ and
$\{a_N,\lambda_N,e_N,\pomega_N\}$, defined in the usual way.  (The
mass of the cental body that enters into the definitions is $M_P+M_C$,
both for Nix and for Pluto-Charon.)  The equations of
motion---Lagrange's planetary equations---are given by Hamilton's
equations for canonical variables, which we may take to be the
Poincar\'e variables , e.g. $\Lambda_{PC}\equiv M_{PC}\sqrt{a_{PC}},
\Lambda_N\equiv M_N\sqrt{a_N}$, etc. \citep{MD99}.
It remains to express the Hamiltonian in terms of the orbital
elements.  The unperturbed terms are \beqn H_{{\rm
unp},PC}&=&-{M_{PC}\over 2 a_{PC}} \\ H_{{\rm unp},N}&=&-{M_N\over 2
a_N} \eeqn For the perturbed term, we resort to the usual Fourier
expansion into a sum of cosine terms.  Appendix B of \cite{MD99}
tabulates the coefficients for the Fourier expansion of
$1/|\bld{r}-\bld{r'}|$ in terms of the orbital elements of one body at
position $\bld{r}$ and an exterior body at position $\bld{r'}$.  (More
precisely, that Appendix tabulates the direct part of the disturbing
function, ${\cal{R}}_D\equiv a'/|\bld{r}-\bld{r'}|$.)  Since equation
(\ref{eq:hpert}) consists of four terms of this form, we can easily
extract the coefficients from \cite{MD99}.  These coefficients are
functions of $\alpha$, the ratio of semi-major axes of the body at
$\bld{r}$ to the one at $\bld{r'}$.  The coefficient for the term ${1/
|\bld{r}_N-(1-\mu_C)\bld{r}_{PC}|}$ should be evaluated at $\alpha =
(1-\mu_C)\alpha_{PC,N}$, where
\be
\alpha_{PC,N}\equiv {a_{PC}\over a_N} \ ;
\ee
similarly, the coefficient $1/r_N$ should be evaluated at $\alpha=0$,
and that for $1/|\bld{r}_N+\mu_C\bld{r}_{PC}|$ at
$\alpha=\mu_C\alpha_{PC,N}$.  But there is an extra complication with
the latter term, since $\mu_C\bld{r}_{PC}$ enters with a positive
sign, instead of a negative one.  To correct for this, one must set
$\pomega_{PC}\rightarrow \pomega_{PC}+\pi$ and
$\lambda_{PC}\rightarrow \lambda_{PC}+\pi$ in the argument of the
cosine; equivalently, one must multiply by $-1$ if within the cosine
argument the sum of the integer coefficients of $\pomega_{PC}$ and of
$\lambda_{PC}$ give an odd number.

The procedure described above yields the equations of motion as
precisely as desired when enough Fourier terms are retained in the
disturbing function $H_{\rm pert}$ (aside from the factor
$M_N/(M_P+M_C)$ dropped from the Hamiltonian.)  But for the purposes
of this paper, it suffices to obtain coefficients that are incorrect
by $\sim \mu_C\sim 10\%$. Therefore we expand $H_{\rm pert}$ to
leading order in $\mu_C$. For each cosine term of the direct potential
${\cal R}_D\equiv a_N/|\bld{r}_N-\bld{r}_{PC}|$ of the form
\be
\hat{\cal R}_D(\alpha_{PC,N})\cos \phi
\ee
 (suppressing
the other arguments of $\hat{\cal R}_D$), 
we have
\be
H_{\rm pert}=-\mu_C{M_N\over a_N}
\hat{\cal R}
\cos\phi
\ ,
\ee
where
\be
\hat{\cal R}\equiv
\hat{\cal R}_D(\alpha_{PC,N})-\hat{\cal R}_D(0)\pm \alpha_{PC,N}({D\hat{\cal R}_D})\vert_{\alpha=0} \ ,
\label{eq:df}
\ee
with $ D\equiv d/d\alpha $ and $\pm \rightarrow +$ if the integer
$d\phi/d\lambda_{PC}+d\phi/d\pomega_{PC}$ is even; otherwise
$\pm\rightarrow -$.

\section{Appendix B. Effect of the 3:1 Resonance on the Secular Damping Rate}

We solve the equations of motion for $z_C$ and $z_N$, including
secular interactions, the 3:1 resonance, as well as tidal damping. The
equations are given in \S \ref{subsec:not}, except here we discard the
term $H_{2:1,CN}$; explicitly,
\beqn
{d\over dt}
\left(\begin{array}{c}z_\C \\z_N\end{array}\right)
&=&i\omn 
\left(\begin{array}{cc}\delta  & -\beta\delta \\ -\beta & 1\end{array}\right)
\left(\begin{array}{c}z_\C \\z_N\end{array}\right)
+ i\nu
\left(\begin{array}{cc}\upsilon\delta & -\chi\delta \\ -\chi & 1\end{array}\right)
\left(\begin{array}{c}z_C^* \\ z_N^* \end{array}\right)e^{i(3\lambda_N-\lambda_C)}
-\gamma_C\left(\begin{array}{c}z_C \\ 0 \end{array}\right) \ ,
\label{eq:3to1eq}
\eeqn
where
\beqn
\nu\equiv 2n_N\mu_Cg_7 \ , \ \ 
\chi \equiv  -{g_6\over  2g_7} \ , \ \ 
\upsilon \equiv {g_5\over g_7} \ ,
\eeqn
and the other quantities are the same as in equation (\ref{eq:zdot}).
To solve these equations analytically, we take the semimajor axes to
be fixed, setting
\beqn
3\lambda_N-\lambda_C=n_{3:1} t
\eeqn
where
\be
n_{3:1}\equiv 3n_N-n_C
\ee
is constant.  We solve the equations perturbatively in the small
parameter $\delta\ll 1$, adopting the same approach as in the
paragraph above equation (\ref{eq:gm}), i.e., we assume that $|z_C|\ll
|z_N|$, which may be verified a postiori.\footnote{If initially
$|z_C|\sim |z_N|$, then $z_C$ would quickly decay on timescale
$\gamma_C^{-1}$ to a value $\sim \delta |z_N|$.}  To leading order in
$|z_C|$,
\be
{dz_N\over dt}\simeq i\eta z_N +i\nu z_N^*e^{in_{3:1}t}  \ . \label{eq:zn1eq}
\ee
Replacing the $\simeq$ with $=$, this equation has solution
\be
z_N = ke^{i(b+\eta)t}+{b\over \nu}k^*e^{i(n_{3:1}-b-\eta)t} \ ,
\label{eq:zn1}
\ee
where $k$ is the complex integration constant, and $b$ is either of the two roots of the
quadratic equation that results from
\be
{b\over \nu}={\nu\over n_{3:1}-2\eta-b} \ .
\ee
For definiteness, we choose the low-frequency root
\be
b={n_{3:1}-2\eta\over 2}+ \left(
\left(n_{3:1}-2\eta\over 2\right)^2-\nu^2
\right)^{1/2} \ .
\ee
In the vicinity of Nix's current semimajor axis $n_{3:1}<0$, implying
that $b\simeq \nu^2/n_{3:1}$ to leading order in $|\eta/n_{3:1}|\sim
2\%$ and $|\nu/n_{3:1}|\sim 4\%$.

Inserting equation (\ref{eq:zn1}) into the approximate equation for $z_C$,
\be
{dz_C\over dt}=-i\eta\beta\delta z_N-i\nu\chi\delta z_N^*e^{in_{3:1}t}-\gamma_Cz_N
\ee
yields
\beqn
z_C=-\delta{\eta\beta+b\chi\over b+\eta-i\gamma_C}ke^{i(b+\eta)t} -
\delta{
\eta\beta b/\nu +\nu\chi
\over n_{3:1}-b-\eta-i\gamma_C}k^*e^{i(n_{3:1}-b-\eta)t} \ ,
\label{eq:zc2}
\eeqn
discarding the homogeneous solution since it decays away after
time $\sim \gamma_C^{-1}$.

Next, we rewrite the full equation for $z_N$ by inserting the approximate solution  (\ref{eq:zn1}) 
into equation (\ref{eq:3to1eq}), with $k$ now time-varying, resulting in  
\be
{dk\over dt}=-i{e^{-i(b+\eta)t}\over 1-b^2/\nu^2}\left[
(\eta\beta+b\chi)z_C+(\eta\beta b/\nu+\nu\chi)z_C^*e^{in_{3:1}t}
\right]
\ee
Upon substitution of equation (\ref{eq:zc2}), we arrive at
\be
{dk\over dt}=ipk + iqk^*e^{i(n_{3:1}-2b-2\eta)t} \ ,
\label{eq:kdot}
\ee
where
\be
p\equiv {\delta\over 1-b^2/\nu^2}\left[ {(\eta \beta+b\chi)^2\over
b+\eta-i\gamma_C} + {(\eta\beta b/\nu+\nu\chi)^2\over
n_{3:1}-b-\eta+i\gamma_C}
\right] \  
\ee
and $q$ is a complex constant whose explicit form we do not give
because we shall have no use for it.  Equation (\ref{eq:kdot}) has the
same form as equation (\ref{eq:zn1eq}), except that now the
coefficients are complex. Its solution has the same form as equation
(\ref{eq:zn1}), except now multiplied by the prefactor $e^{-[{\rm
Im}(p)]t}$.  In conclusion, on timescales much longer than
$1/\gamma_C$, $z_N$ slowly decays at the rate $\gamma_{3:1}\equiv {\rm
Im}(p)$, or
\be
\gamma_{3:1} = {\delta\gamma_C \over 1-b^2/\nu^2}\left[
\left(
{\eta\beta+b\chi\over b+\eta}
\right)^2
-
\left(
{\eta\beta b/\nu+\nu\chi\over n_{3:1}-b-\eta}
\right)^2
\right] \ .
\label{eq:gamma31}
\ee
This rate is plotted in Figure \ref{fig:3to1} relative to damping rate
in the absence of the 3:1 forcing term.

\section{Appendix C: Hydra}
\label{sec:hydra}

Hydra's evolution in the presence of Chaon is similar to Nix's, but
with some quantitative differences.  Hydra's purely secular damping
rate relative to Nix's is (eq. [\ref{eq:gpm}])
\be
{\gamma_{-H}\over \gamma_{-N}}={M_H\over M_N}{ \sqrt{a_H}\over\sqrt{a_N}}{\beta_H^2\over\beta_N^2}
= 0.7 {M_H\over M_N} \ ,
\ee
where subscripts N and H are for Nix and Hydra, and the numerical
expression is evaluated for Nix and Hydra at, respectively, the
nominal 4:1 and 6:1 resonances with Charon.  The 3:1 resonance has a
relatively small effect on Hydra's secular damping rate: from the
extrapolation of Figure \ref{fig:3to1} to the nominal 6:1 resonance,
it reduces $\gamma_{-H}$ by the factor $0.6$.  Hydra's forced
eccentricity due to the 2:1 resonance with Charon is smaller than
Nix's by (eq. [\ref{eq:zn21}])
\be
\left|{z_{H,2:1}\over z_{N,2:1}}\right|={g_{4,H}\over g_{4,N}}{n_H\over n_N}{n_{2:1,N}\over n_{2:1,H}}
= 0.2  \ ,
\ee
and its migration rate is reduced by (eq. [\ref{eq:kap}])
\be
{\kappa_H\over \kappa_N}= {M_H\sqrt{a_H}\over M_N\sqrt{a_N}}
\left({g_{3,H}\over g_{3,N}}
{n_H\over n_N}{n_{2:1,N}\over n_{2:1,H}}
\right)^2= 0.09 {M_H\over M_N}
\ee

It is also interesting to consider the interaction between Nix and
Hydra.  Hydra's reported eccentricity based on a Keplerian fit differs from zero,
$e_H=0.0052(11)$, and this might be due to the proximity of Nix and
Hydra to their mutual 3:2 resonance.  The 3:2 interaction energy
between Nix and Hydra is
\be
H_{3:2,NH}=-\mu_N{M_H\over 2a_H}\left(
g_az_N^*+g_bz_H^*
\right)e^{i(3\lambda_H-2\lambda_N)} + c.c. \ ,
\ee
where $\mu_N\equiv M_N/(M_C+M_P)$, $g_a=-(6+\alpha
D)b_{1/2}^3/2=-2.03$, and $g_b=(5+\alpha D)b_{1/2}^2/2=2.48$, using
$\alpha=(2/3)^{2/3}$ in the numerical expressions.  Using the equation
for Hydra's eccentricity, it is simple to show that the 3:2 resonant
term gives a forced eccentricity to Hydra equal to
\be
z_{H,3:2}=\mu_N g_b {n_H\over
3n_H-2n_N-\eta_H}e^{i(3\lambda_H-2\lambda_N)} .
\ee
With the observationally derived values for $n_H$ and $n_N$ given in
the Introduction, we find $|z_{H,3:2}|=0.003 {\mu_N/10^{-4}}$, close
to Hydra's reported eccentricity. {Since Nix and Hydra are
close to 3:2 resonance, this eccentricity vector precesses at a slower
rate than other resonantly forced eccentricity vectors (which precess
at orbital timescales). It will show up in a Keplerian fit that covers
many orbital periods. And we suggest that this likely explains the
observed eccentricity of Hydra. Nix's 3:2 forced eccentricity has a
similar expression but scales with $\mu_H$. \citet{TBGE07} report a
Hydra mass that is lower by a factor of 2 than Nix and encompasses
zero. This is consistent with Nix's lower measured eccentricity.}

\bibliographystyle{apj}
\bibliography{plutoy}

\end{document}